\documentclass[journal]{IEEEtran}
\usepackage[linesnumbered,ruled]{algorithm2e}
\usepackage{graphicx,amssymb,mathrsfs,amsmath,array,color}
\usepackage{subfigure}
\usepackage{multirow}
\usepackage{booktabs}
\usepackage{cite}
\usepackage[colorlinks,urlcolor=blue,driverfallback=dvipdfm]{hyperref}

\newcommand{\norm}[1]{\lVert#1\rVert}
\DeclareMathOperator{\F}{F}

\begin{document}
\title{Spectral Superresolution of Multispectral Imagery with Joint Sparse and Low-Rank Learning}

\author{Lianru Gao,~\IEEEmembership{Senior Member,~IEEE,}
        Danfeng Hong,~\IEEEmembership{Member,~IEEE,}
        Jing Yao,
        Bing Zhang,~\IEEEmembership{Fellow,~IEEE,}
        Paolo Gamba,~\IEEEmembership{Fellow,~IEEE,}
        and Jocelyn Chanussot,~\IEEEmembership{Fellow,~IEEE}

\thanks{This work was supported by the National Natural Science Foundation of China under Grant 41722108 and Grant 91638201 as well as with the support of the AXA Research Fund. (\emph{Corresponding author: Danfeng Hong}).}
\thanks{L. Gao is with the Key Laboratory of Digital Earth Science, Aerospace Information Research Institute, Chinese Academy of Sciences, Beijing 100094, China. (e-mail: gaolr@aircas.ac.cn)}
\thanks{D. Hong is with the Univ. Grenoble Alpes, CNRS, Grenoble INP, GIPSA-lab, Grenoble 38000, France. (e-mail: hongdanfeng1989@gmail.com)}
\thanks{J. Yao is with the School of Mathematics and Statistics, Xi’an Jiaotong University, 710049 Xi’an, China. (e-mail: jasonyao@stu.xjtu.edu.cn)}
\thanks{B. Zhang is with the Key Laboratory of Digital Earth Science, Aerospace Information Research Institute, Chinese Academy of Sciences, Beijing 100094, China, and also with the College of Resources and Environment, University of Chinese Academy of Sciences, Beijing 100049, China. (e-mail: zb@radi.ac.cn)}
\thanks{P. Gamba is with Department of Electrical, Computer and Biomedical Engineering, University of Pavia, Pavia 27100, Italy. (e-mail:paolo.gamba@unipv.it)}
\thanks{J. Chanussot is with the Univ. Grenoble Alpes, INRIA, CNRS, Grenoble INP, LJK, Grenoble 38000, France, and with the Key Laboratory of Digital Earth Science, Aerospace Information Research Institute, Chinese Academy of Sciences, Beijing 100094, China. (e-mail: jocelyn@hi.is)}
}

\markboth{Submission to IEEE Transactions on Geoscience and Remote Sensing,~Vol.~XX, No.~XX, ~XXXX,~2020}
{Shell \MakeLowercase{\textit{et al.}}: Spectral Superresolution of Multispectral Image with Joint Sparse Learning}

\maketitle
\begin{abstract}
Extensive attention has been widely paid to enhance the spatial resolution of hyperspectral (HS) images with the aid of multispectral (MS) images in remote sensing. However, the ability in the fusion of HS and MS images remains to be improved, particularly in large-scale scenes, due to the limited acquisition of HS images. Alternatively, we super-resolve MS images in the spectral domain by the means of partially overlapped HS images, yielding a novel and promising topic: spectral superresolution (SSR) of MS imagery. This is challenging and less investigated task due to its high ill-posedness in inverse imaging. To this end, we develop a simple but effective method, called \underline{j}oint \underline{s}parse and \underline{lo}w-rank \underline{l}earning (J-SLoL), to spectrally enhance MS images by jointly learning low-rank HS-MS dictionary pairs from overlapped regions. J-SLoL infers and recovers the unknown hyperspectral signals over a larger coverage by sparse coding on the learned dictionary pair. Furthermore, we validate the SSR performance on three HS-MS datasets (two for classification and one for unmixing) in terms of reconstruction, classification, and unmixing by comparing with several existing state-of-the-art baselines, showing the effectiveness and superiority of the proposed J-SLoL algorithm. Furthermore, the codes and datasets will be available at: \url{https://github.com/danfenghong/IEEE\_TGRS\_J-SLoL}, contributing to the RS community.
\end{abstract}
\graphicspath{{figures/}}


\begin{figure}[!t]
	  \centering
		\subfigure{
			\includegraphics[width=0.5\textwidth]{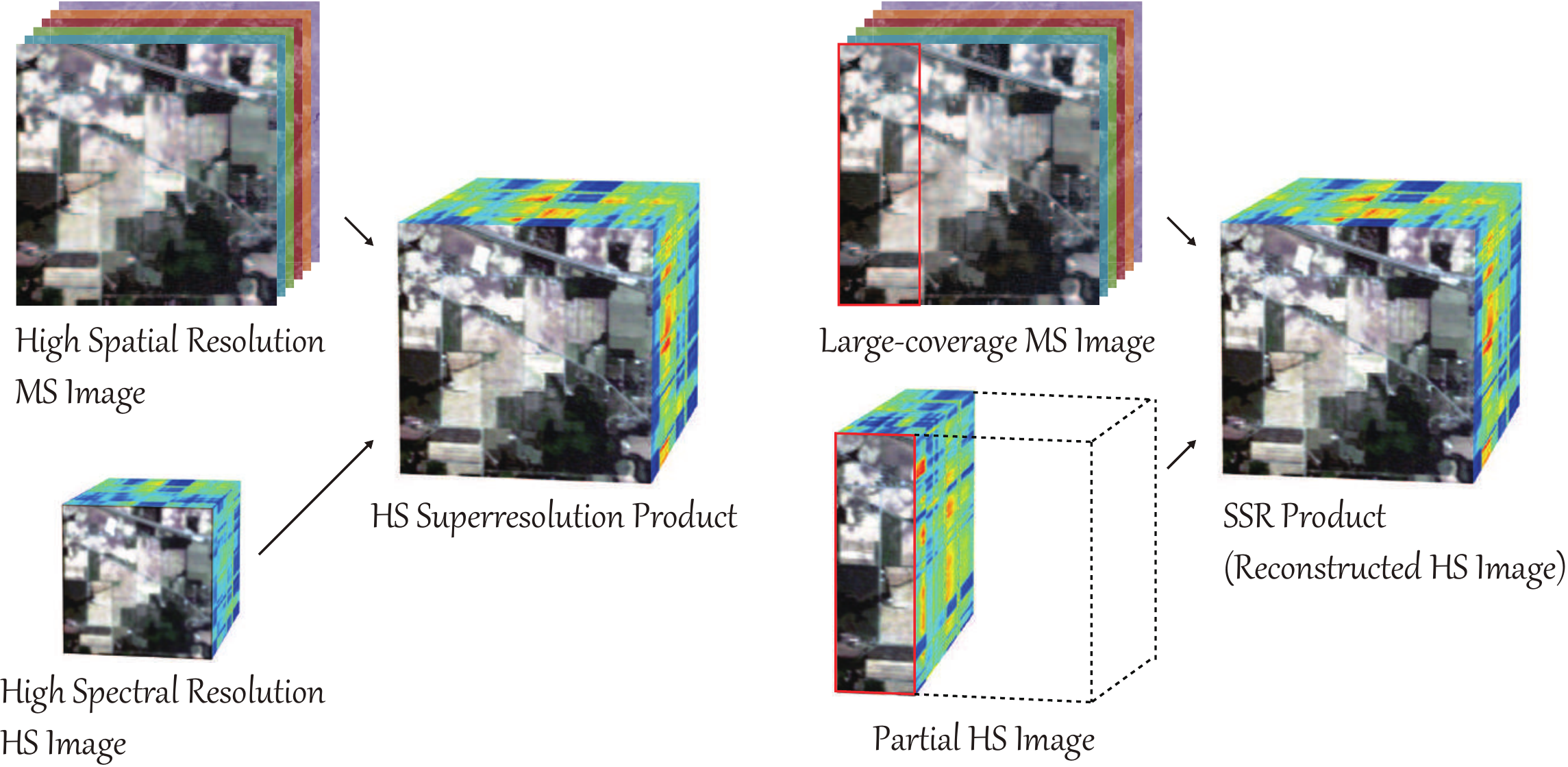}
		}
        \caption{Illustrative comparison for the tasks of HS-SR and SSR of MS images.}
\label{fig:motivations}
\end{figure}

\section{Introduction}
\IEEEPARstart{W}{ith} the rapid development and enormous breakthroughs in imaging technology, a variety of image products, e.g., hyperspectral (HS) data, multispectral (MS) data, synthetic aperture radar (SAR), light detection and ranging (LiDAR), have been widely and successfully applied to many practical applications in remote sensing (RS) and geoscience \cite{yuan2020deep, gao2017new, yuan2020estimating}, such as mineral exploration, urban planning and management, previous framing, water quality assessment, disaster pre-warning and prevention, to name a few.

Characterized by very rich and diverse spectral information, HS images, which enable to identify the materials on the surface of the Earth more accurately, have been paid an increasing attention in various HS RS tasks: feature extraction and embedding \cite{ gao2017optimized,hong2017learning,yao2018robust,hong2018joint,licciardi2018spectral,cao2020an,rasti2020feature,cao2020hyperspectral}, spectral unmixing \cite{tang2017integrating,yao2019nonconvex,hong2019augmented}, data fusion \cite{liu2019stfnet,xu2019nonlocal,Hu2019mima,wang2019large, xu2020hyperspectral}, target detection \cite{wu2019fourier,wu2019approximate,wu2019orsim}, and multimodal data analysis \cite{hang2020classification,hong2020learning}. However, although currently operational, e.g., Earth Observing-1 (EO-1), DLR Earth Sensing Imaging Spectrometer (DESIS), or upcoming imaging spectroscopy, e.g., Environmental Mapping and Analysis Program (EnMAP), satellite missions and airborne imaging sensors can provide HS data of high spectral resolution, yet its low spatial resolution limits the performance to be further improved. 

\begin{figure*}[!t]
	  \centering
		\subfigure{
			\includegraphics[width=0.9\textwidth]{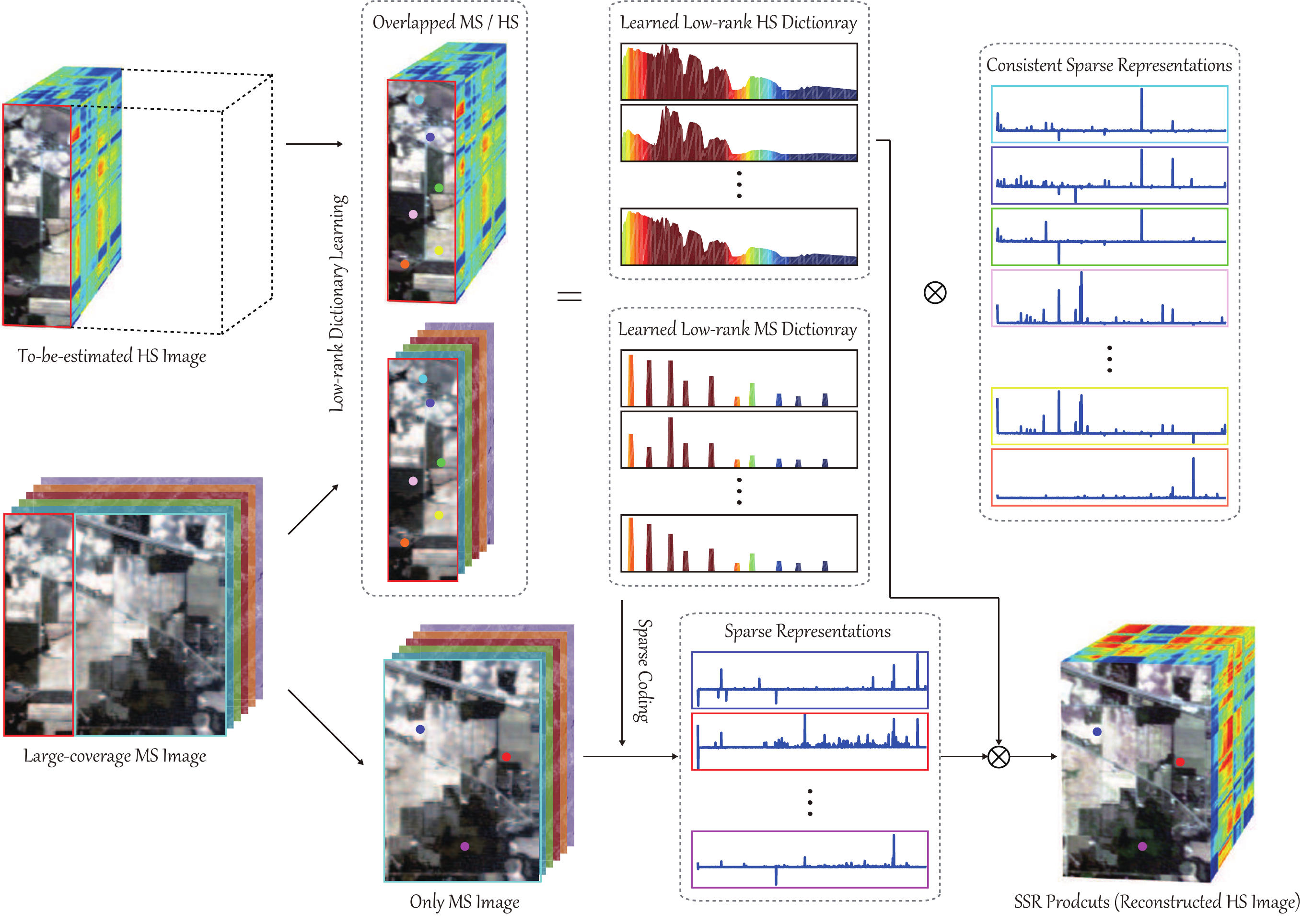}
		}
        \caption{An illustration of the proposed J-SLoL model for the SSR task. It consists of two parts: low-rank dictionary learning (learned HS-MS dictionary pair) and sparse coding for the final SSR product (reconstructed HS image).}
\label{fig:workflow}
\end{figure*}

For this reason, enormous effects have been recently made to enhance the spatial resolution of HS images by fusing corresponding MS images \cite{yokoya2017hyperspectral}, yielding high-quality HS products (high spatial and spectral resolutions). Yokoya \textit{et al.} \cite{yokoya2011coupled} developed a classic but very effective work for the fusion of HS and MS images by coupled spectral unmixing. Such a strategy has been proven to be effective by many follow-up works. For example, authors in \cite{lanaras2015hyperspectral} re-formulated the coupled spectral unmixing model as a constrained optimization problem for HS superresolution (HS-SR). Beyond the pixel-based fusion, Xu \textit{et al.} \cite{xu2019nonlocal} proposed a nonlocal coupled spectral unmixing in a tensorized manner by considering a larger perspective filed of each pixel to reconstruct the fused HS image. The same investigators \cite{xu2020hyperspectral_TNN} further extended their work by adaptively learning response functions instead of pre-given ones. Besides, there are many other types of HS-MS fusion methods, e.g., dictionary learning \cite{akhtar2014sparse}, sparse representation \cite{wei2015hyperspectral}, Bayesian fusion \cite{wei2015bayesian}, subspace learning \cite{hong2019cospace,xu2020hyperspectral}, deep learning-based approaches \cite{xie2019multispectral}, etc.

It is well known, however, that the HS data fail to be acquired in a large-covered area due to the limitations of imaging devices. This would bring a big challenge in collecting the same-size HS-MS image pairs for the fusion task. But fortunately, we may expect to have the MS data \cite{hong2015novel} at a large and even global scale, owing to its easy-availability. We have to admit that the relatively poor spectral information makes difficulty for the MS data to detect and recognize the materials at a more accurate level, particularly for those classes that hold very similar visual cues. As a trade-off, spectral superresolution (SSR) of MS images might be an alternative solution, as illustrated in Fig. \ref{fig:motivations}. This topic is novel and promising. Although some famous works \cite{arad2016sparse,aeschbacher2017defense,galliani2017learned,can2018efficient,akhtar2018hyperspectral} have been studied by the attempts to challenge the similar topic from the perspective of methodology in the computer vision field, yet it is less investigated by researchers in RS. There are only several tentative works. For example, authors of \cite{winter2007hyperspectral} simply assumed the existence of a regression matrix between overlapped HS and MS regions and utilized the estimated transformation to reconstruct the unknown HS signals. Further, Sun \textit{et al.} \cite{sun2014enhancement} learned multiple transformation matrices from the grouped HS image and screened out an optimal one by a weighted spectral angle distance. Another recent work related to this task was presented in \cite{yokoya2018spectral}, in which sparse representation is used to recover the large-scale HS image from partially overlapped HS and MS images. It should be noted that without any priors (or regularizations), estimating simple regression matrix from the limited HS-MS pairs (e.g., \cite{winter2007hyperspectral} and \cite{sun2014enhancement}) is a highly ill-conditioned problem, thereby leading to poor reconstruction for HS images. On the other hand, sparse representation in \cite{yokoya2018spectral} is capable of recovering the unknown HS signals well. Nevertheless, reconstruction coefficients (or sparse representations) are only estimated on the MS data taken as the dictionary (or basis) and directly transferred into the HS reconstruction. Consequently, the two kinds of approaches could be effective for the SSR task to some extent, but the potential in fully making use of overlapped HS-MS images remains limited.

To overcome these difficulties, we propose a simple but effective learning algorithm, called \textbf{j}oint \textbf{s}parse and \textbf{lo}w-rank \textbf{l}earning (J-SLoL), for the task of SSR. As the name suggests, J-SLoL jointly learns the low-rank HS-MS dictionary pair and its corresponding sparse representations. The learned HS and MS dictionaries are consistent in the activated locations of sparse coefficients for each pixel. Such consistency enables the reconstruction of HS images at a more accurate level. More specifically, main contributions of this paper can be highlighted as follows

\begin{itemize}
    \item We equivalently convert the problem of HS-SR to that of the SSR of MS images. Compared to the former, the latter has great potential in recovering high-quality HS images, particularly in large-coverage case. To our best knowledge, this is the first time to investigate the SSR problem in RS.
    \item We propose a J-SLoL model for addressing the SSR problem effectively. J-SLoL can learn low-rank overcomplete dictionaries with respect to HS and MS data, respectively, and consistent sparse representations from the overlapped HS-MS images and further reconstruct the unknown HS images by using shared sparse coefficients obtained from the corresponding MS parts.
    \item Reconstruction, classification, and unmixing are used to evaluate the product quality. Extensive experiments conducted on three HS-MS datasets demonstrate the superiority of the proposed J-SLoL model in the SSR case.
\end{itemize}

The remaining part of this paper is organized as follows. We briefly state the SSR problem formulation and present the proposed J-SLoL model and its optimizer in Section II. Section III provides extensive experiments as well as corresponding results and analysis. Finally, we make a conclusion to this paper with a future outlook in Section IV.

\section{Joint Sparse and Low-Rank Learning}
\subsection{Problem Statement}
It is clear that our goal is to obtain the HS product of high spatial and high spectral resolutions. Therefore, HS-SR and SSR of MS images are two feasible solutions, as shown in Fig. \ref{fig:motivations}. It should be noted, however, that the main advantage of SSR using MS images over HS-SR lies in easy availability of MS images on a larger geospatial coverage. On the other hand, the HS imagery is generally acquired in the form of a very narrow strip, as the spectrum within one pixel consists of hundreds of wavelength bands, limiting the imaging range of HS images. 

Unlike the HS-SR task that the same geospatial region for HS and MS images is needed, SSR aims to enhance spectral resolution of MS images only using partially overlapped HS and MS images (see Fig. \ref{fig:motivations}). This can save the cost in data preparation well and enables the generation of high-quality HS images in an easier way. Please note that a simplified case of SSR in this paper is investigated by using partially HS-MS images at a same ground sampling distance (GSD).

\subsection{Problem Formulation}
Fig. \ref{fig:workflow} illustrates the proposed J-SLoL model, where we reasonably assume each spectrum out of the overlapped region, either in the HS or the MS images, can be well reconstructed by an identical sparse combination of atoms on the low-rank HS (or MS) dictionary learned from the overlapped HS and MS images.

Let the spectral signatures of the HS and MS images in the overlapped part be $\mathbf{H}_{in}=[\mathbf{h}_{in}^{1},...,\mathbf{h}_{in}^{i},...,\mathbf{h}_{in}^{N}]\in \mathbb{R}^{P\times N}$ with $P$ spectral bands by $N$ pixels and $\mathbf{M}_{in}=[\mathbf{m}_{in}^{1},...,\mathbf{m}_{in}^{i},...,\mathbf{m}_{in}^{N}]\in \mathbb{R}^{Q\times N}$ with $Q$ channels by $N$ pixels, respectively. Moreover, we define those pixels out of the overlapped region as $\mathbf{H}_{out}=[\mathbf{h}_{out}^{1},...,\mathbf{h}_{out}^{j},...,\mathbf{h}_{out}^{N_1}]\in \mathbb{R}^{P\times N_1}$ and $\mathbf{M}_{out}=[\mathbf{m}_{out}^{1},...,\mathbf{m}_{out}^{j},...,\mathbf{m}_{out}^{N_1}]\in \mathbb{R}^{Q\times N_1}$ for the HS and MS images, respectively. Our J-SLoL model consists of two steps: low-rank dictionary learning and sparse recovery, for the SSR task. 

\subsubsection{Low-Rank Dictionary Learning (D-Step)}
This process can be formulated by solving the following constrained optimization problem
\begin{equation}
\label{eq1}
\begin{aligned}
       \mathop{\min}_{\mathbf{D}_{h},\mathbf{D}_{m},\mathbf{X}}&\frac{1}{2}\norm{\mathbf{H}_{in}-\mathbf{D}_{h}\mathbf{X}}_{\F}^{2}+\frac{\alpha}{2}\norm{\mathbf{M}_{in}-\mathbf{D}_{m}\mathbf{X}}_{\F}^{2}+\beta\norm{\mathbf{X}}_{1,1}\\
       &+\gamma(\norm{\mathbf{D}_{h}}_{*}+\norm{\mathbf{D}_{m}}_{*}), \\
       &\mathrm{s.t.} \; \; \mathbf{D}_{h} \succeq \mathbf{0}, \; \; \mathbf{D}_{m}\succeq \mathbf{0}, \;\; \mathbf{1}^\top\mathbf{X}=\mathbf{1},
\end{aligned}
\end{equation}
where $\mathbf{D}_{h}\in \mathbb{R}^{P\times L}$ and $\mathbf{D}_{m}\in \mathbb{R}^{Q\times L}$ are the to-be-learned low-rank dictionaries with respect to HS and MS data, and $\mathbf{X}\in \mathbb{R}^{L\times N}$ denotes the consistent sparse representations on the two dictionaries. $\norm{\bullet}_{*}$ and $\norm{\bullet}_{1,1}$ represent the nuclear norm to approximate the rank of matrix and the sparsity-prompting term approximately estimated by $\ell_{1}$-norm, respectively. Moreover, $\mathbf{1}$ and $\mathbf{0}$ are the unit vector and the zero matrix, respectively. $\alpha$, $\beta$, and $\gamma$ are penalty parameters to balance the importance of different terms in Eq. (\ref{eq1}).

\subsubsection{Sparse Recovery (S-Step)}
Owing to the consistent sparse representations in dictionary learning, the learned HS and MS dictionaries are well applicable to the unknown HS reconstruction. Therefore, there will be two main parts in sparse recovery, i.e., sparse coding on $\mathbf{D}_{m}$ and HS reconstruction using $\mathbf{D}_{h}$.

The sparse coefficients of $\mathbf{M}_{out}$ can be encoded on the MS dictionary ($\mathbf{D}_{m}$), the resulting sparse coding problem can be then written as
\begin{equation}
\label{eq2}
\begin{aligned}
       \mathop{\min}_{\mathbf{Y}}\frac{1}{2}\norm{\mathbf{M}_{out}-\mathbf{D}_{m}\mathbf{Y}}_{\F}^{2}+\eta\norm{\mathbf{Y}}_{1,1}, \;\;\; \mathrm{s.t.} \; \; \mathbf{1}^\top\mathbf{Y}=\mathbf{1},
\end{aligned}
\end{equation}
where $\mathbf{Y}$ denotes the sparse coefficients with respect to the variable $\mathbf{M}_{out}$.

Once the coding results ($\mathbf{Y}$) are given, it is straightforward to derive the HS reconstruction (or SSR of MS images), denoted as $\mathbf{\hat{H}}_{out}=\mathbf{D}_{h}\mathbf{Y}$.

\subsection{Model Optimization}
The proposed J-SLoL model consists of two optimization problems: low-rank dictionary learning and sparse coding, where the latter is a special case of the former. 

\subsubsection{D-Step Solver}
Despite the non-convexity in Eq. (\ref{eq1}), the sub-problems for each variable is solvable. For that, an alternating direction method of multipliers (ADMM) algorithm is designed to solve this model fast and effectively. To facilitate the use of ADMM optimizer, an equivalent form of Eq. (\ref{eq1}) is converted by introducing several auxiliary variables, i.e., $\mathbf{Z}$, $\mathbf{J}$, and $\mathbf{K}$, to replace the to-be-estimated variables $\mathbf{X}$, $\mathbf{D}_{h}$, and $\mathbf{D}_{m}$, respectively, we then have
\begin{equation}
\label{eq3}
\begin{aligned}
       \mathop{\min}_{\mathcal{S}}&\frac{1}{2}\norm{\mathbf{H}_{in}-\mathbf{D}_{h}\mathbf{X}}_{\F}^{2}+\frac{\alpha}{2}\norm{\mathbf{M}_{in}-\mathbf{D}_{m}\mathbf{X}}_{\F}^{2}+\beta\norm{\mathbf{Z}}_{1,1}\\
       &+\gamma(\norm{\mathbf{J}}_{*}+\norm{\mathbf{K}}_{*})+l_R^+(\mathbf{J})+l_R^+(\mathbf{K}), \\
       &\mathrm{s.t.} \; \; \mathbf{D}_{h} = \mathbf{J}, \;\; \mathbf{D}_{m} = \mathbf{K}, \;\; \mathbf{X} = \mathbf{Z}, \;\; \mathbf{1}^\top\mathbf{X}=\mathbf{1},
\end{aligned}
\end{equation}
where the variable set $\mathcal{S}=\{\mathbf{X},\mathbf{D}_{h},\mathbf{D}_{m},\mathbf{Z}, \mathbf{J},\mathbf{K}\}$, and the symbol $()^{+}$ is an element-wise positive operator that truncates the non-negative part of the vector or matrix, i.e., $l_{R}^{+}(\bullet)$ means $\bullet \succeq \mathbf{0}$. Further, the augmented Lagrangian function of problem (\ref{eq3}) $\mathcal{L}$ can be equivalently written in the form of
\begin{equation}
\label{eq4}
\begin{aligned}
       \mathcal{L}(\mathcal{S},& \{\mathbf{\Lambda}_{i}\}_{i=1}^{3})=\frac{1}{2}\norm{\mathbf{H}_{in}-\mathbf{D}_{h}\mathbf{X}}_{\F}^{2}+\frac{\alpha}{2}\norm{\mathbf{M}_{in}-\mathbf{D}_{m}\mathbf{X}}_{\F}^{2}\\
       &+\beta\norm{\mathbf{Z}}_{1,1}+\gamma(\norm{\mathbf{J}}_{*}+\norm{\mathbf{K}}_{*})+l_R^+(\mathbf{J})+l_R^+(\mathbf{K})\\
       &+\mathbf{\Lambda}_{1}^\top(\mathbf{Z}-\mathbf{X})+\mathbf{\Lambda}_{2}^\top(\mathbf{J}-\mathbf{D}_{h})+\mathbf{\Lambda}_{3}^\top(\mathbf{K}-\mathbf{D}_{m})\\
       &+\frac{\mu}{2}\norm{\mathbf{Z}-\mathbf{X}}_{\F}^{2}+\frac{\mu}{2}\norm{\mathbf{J}-\mathbf{D}_{h}}_{\F}^{2}+\frac{\mu}{2}\norm{\mathbf{K}-\mathbf{D}_{m}}_{\F}^{2},\\
       &\mathrm{s.t.} \; \; \mathbf{1}^\top\mathbf{X}=\mathbf{1},
\end{aligned}
\end{equation}
where $\{\mathbf{\Lambda}_{i}\}_{i=1}^{3}$ and $\mu$ are the Lagrange multipliers and the positive penalty parameter, respectively. These variables ($\mathcal{S}$ and $\{\mathbf{\Lambda}_{i}\}_{i=1}^{3}$) in $\mathcal{L}$ can be successively minimized as follows:

\begin{algorithm}[!t]
\caption{J-SLoL solver: \textit{D-Step}}
\KwIn{$\mathbf{H}_{in},\mathbf{M}_{in},$ parameters $\alpha,\beta,$ and $\mathrm{maxIter}.$}
\KwOut{$\mathbf{D}_{h},\mathbf{D}_{m},$ and $\mathbf{X}.$}
\textbf{Initialization}: $\mathbf{J}^{0}=\mathbf{0}, \mathbf{K}^{0}=\mathbf{0}, \mathbf{Z}^{0}=\mathbf{0}, \{\mathbf{\Lambda}_{i}^{0}\}_{i=1}^{3}=\mathbf{0}, \mu^{0}=10^{-3}, \mu_{max}=10^{6}, \xi=1.5, \varepsilon=10^{-6}, t=1.$\\
  \While{not converged \rm{or} $t>\mathrm{maxIter}$}
 {
         Update the variable $\mathbf{X}^{t}$ by Eq. (\ref{eq6});\\
         Update the variable $\mathbf{D}_{h}^{t}$ by Eq. (\ref{eq8});\\
         Update the variable $\mathbf{D}_{m}^{t}$ by Eq. (\ref{eq10});\\
         Update the variables $\mathbf{J}^{t}$ and $\mathbf{K}^{t+1}$ by Eq. (\ref{eq13});\\
         Update the variable $\mathbf{Z}^{t}$ by Eq. (\ref{eq15});\\
         Update Lagrange multipliers $\{\mathbf{\Lambda}_{i}^{t}\}_{i=1}^{3}$ by Eq. (\ref{eq16});\\
         Update the parameter $\mu^{t}$ by
         $\min (\xi\mu^{t-1},\mu_{max})$;\\
         Check the convergence conditions:\\
         \eIf{$\norm {\mathbf{Z}^{t}-\mathbf{X}^{t}}_{\F}<\varepsilon$ and $\norm {\mathbf{J}^{t}-\mathbf{D}_{h}^{t}}_{\F}<\varepsilon$ and $\norm {\mathbf{K}^{t}-\mathbf{D}_{m}^{t}}_{\F}<\varepsilon$}
         {
           Stop iteration;
         }
         {
           $t\leftarrow t+1$;
         }
 }
\end{algorithm}

\textit{Optimization with respect to $\mathbf{X}$}: We solve the following constrained optimization problem for the variable $\mathbf{X}$:
\begin{equation}
\label{eq5}
\begin{aligned}
       \mathop{\min}_{\mathbf{X}}&\frac{1}{2}\norm{\mathbf{H}_{in}-\mathbf{D}_{h}\mathbf{X}}_{\F}^{2}+\frac{\alpha}{2}\norm{\mathbf{M}_{in}-\mathbf{D}_{m}\mathbf{X}}_{\F}^{2}\\
       &+\mathbf{\Lambda}_{1}^\top(\mathbf{Z}-\mathbf{X})+\frac{\mu}{2}\norm{\mathbf{Z}-\mathbf{X}}_{\F}^{2},\;\;\;\mathrm{s.t.} \; \; \mathbf{1}^\top\mathbf{X}=\mathbf{1}.
\end{aligned}
\end{equation}
The closed-form solution of problem (\ref{eq5}) is given by
\begin{equation}
\label{eq6}
\begin{aligned}
       \mathbf{X} \leftarrow \mathbf{A}^{-1}\mathbf{B}-\mathbf{C}(\mathbf{I}_{1\times L}\mathbf{A}^{-1}\mathbf{B}-1),
\end{aligned}
\end{equation}
where 
\begin{equation}
\nonumber
\begin{aligned}
       &\mathbf{A}=\mathbf{D}_{h}^\top\mathbf{D}_{h}+\alpha\mathbf{D}_{m}^\top\mathbf{D}_{m}+\mu\mathbf{I}_{L\times L},\\
       &\mathbf{B}=\mathbf{D}_{h}^\top\mathbf{H}_{in}+\alpha\mathbf{D}_{m}^\top\mathbf{M}_{in}+\mu\mathbf{Z}+\mathbf{\Lambda}_{1},\\
       &\mathbf{C}=\mathbf{A}^{-1}\mathbf{I}_{L\times 1}(\mathbf{I}_{1\times L}\mathbf{A}^{-1}\mathbf{I}_{L\times 1})^{-1}.
       \end{aligned}
\end{equation}

\begin{algorithm}[!t]
\caption{J-SLoL solver: \textit{S-Step}}
\KwIn{$\mathbf{M}_{out},\mathbf{D}_{h},\mathbf{D}_{m},$ parameters $\eta,$ and $\mathrm{maxIter}.$}
\KwOut{$\mathbf{Y}.$}
\textbf{Initialization}: $\mathbf{O}^{0}=\mathbf{0}, \mathbf{\Delta}^{0}=\mathbf{0}, \rho^{0}=10^{-3}, \rho_{max}=10^{6}, \xi=1.5, \varepsilon=10^{-6}, t=1.$\\
  \While{not converged \rm{or} $t>\mathrm{maxIter}$}
 {
         Update the variable $\mathbf{Y}^{t}$ by Eq. (\ref{eq19});\\
         Update the variable $\mathbf{O}^{t}$ by Eq. (\ref{eq20});\\
         Update Lagrange multiplier $\mathbf{\Delta}^{t}$ by Eq. (\ref{eq21});\\
         Update the parameter $\rho^{t}$ by
         $\min (\xi\rho^{t-1},\rho_{max})$;\\
         Check the convergence conditions:\\
         \eIf{$\norm {\mathbf{O}^{t}-\mathbf{Y}^{t}}_{\F}<\varepsilon$}
         {
           Stop iteration;
         }
         {
           $t\leftarrow t+1$;
         }
 }
\end{algorithm}

\textit{Optimization with respect to $\mathbf{D}_{h}$}: The optimization problem of $\mathbf{D}_{h}$ can be expressed by
\begin{equation}
\label{eq7}
\begin{aligned}
       \mathop{\min}_{\mathbf{D}_{h}}&\frac{1}{2}\norm{\mathbf{H}_{in}-\mathbf{D}_{h}\mathbf{X}}_{\F}^{2}+\mathbf{\Lambda}_{2}^\top(\mathbf{J}-\mathbf{D}_{h})+\frac{\mu}{2}\norm{\mathbf{J}-\mathbf{D}_{h}}_{\F}^{2},
\end{aligned}
\end{equation}
which has the following analytical resolution
\begin{equation}
\label{eq8}
\begin{aligned}
       \mathbf{D}_{h} \leftarrow (\mathbf{H}_{in}\mathbf{X}^\top+\mu\mathbf{J}+\mathbf{\Lambda}_{2})\times (\mathbf{X}\mathbf{X}^\top+\mu\mathbf{I}_{L\times L})^{-1}.
\end{aligned}
\end{equation}

\textit{Optimization with respect to $\mathbf{D}_{m}$}: Similarly to problem (\ref{eq7}), the objective function for the variable $\mathbf{G}_{m}$ is 
\begin{equation}
\label{eq9}
\begin{aligned}
       \mathop{\min}_{\mathbf{D}_{m}}&\frac{\alpha}{2}\norm{\mathbf{M}_{in}-\mathbf{D}_{m}\mathbf{X}}_{\F}^{2}+\mathbf{\Lambda}_{3}^\top(\mathbf{K}-\mathbf{D}_{m})+\frac{\mu}{2}\norm{\mathbf{K}-\mathbf{D}_{m}}_{\F}^{2},
\end{aligned}
\end{equation}
whose solution can be directly derived by
\begin{equation}
\label{eq10}
\begin{aligned}
       \mathbf{D}_{m} \leftarrow (\alpha\mathbf{M}_{in}\mathbf{X}^\top+\mu\mathbf{K}+\mathbf{\Lambda}_{3})\times (\alpha\mathbf{X}\mathbf{X}^\top+\mu\mathbf{I}_{L\times L})^{-1}.
\end{aligned}
\end{equation}

\begin{figure*}[!t]
	  \centering
		\subfigure[Indian Pines Data]{
			\includegraphics[width=0.25\textwidth]{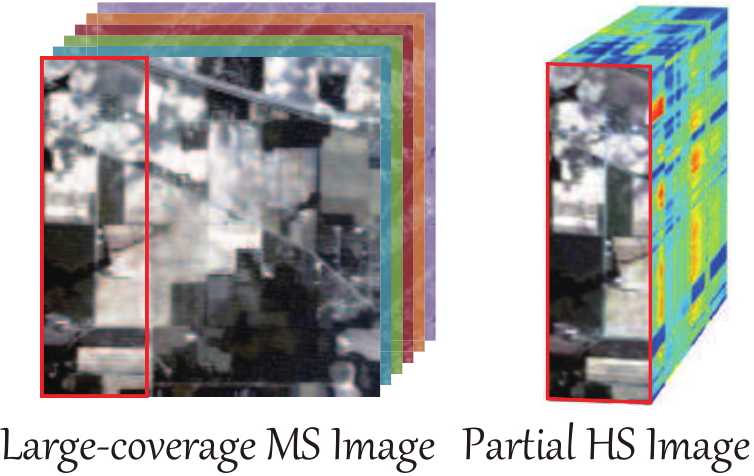}
            \label{fig:IP}
		}\qquad
		\subfigure[Pavia University Data]{
			\includegraphics[width=0.25\textwidth]{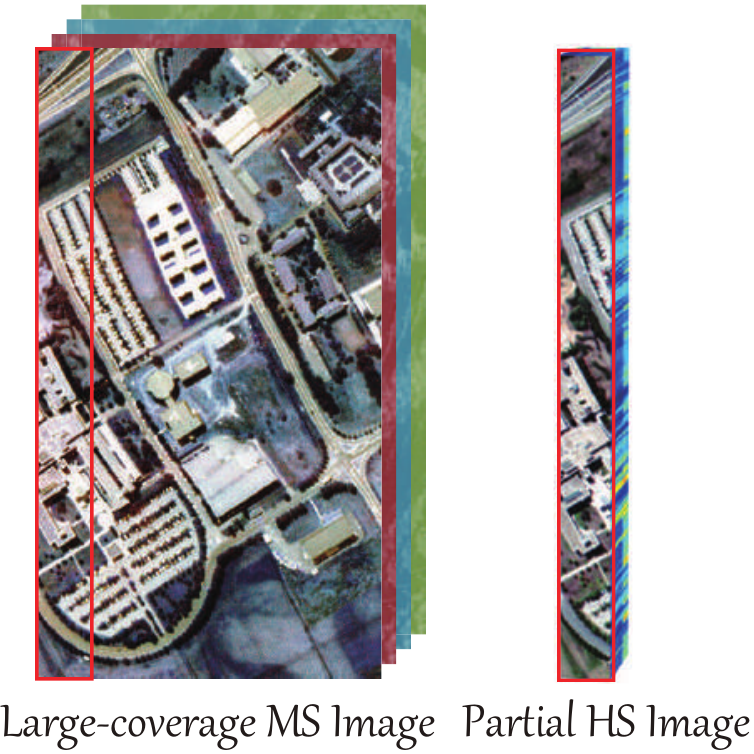}
            \label{fig:PU}
		}\qquad
		\subfigure[Jasper Ridge Data]{
			\includegraphics[width=0.25\textwidth]{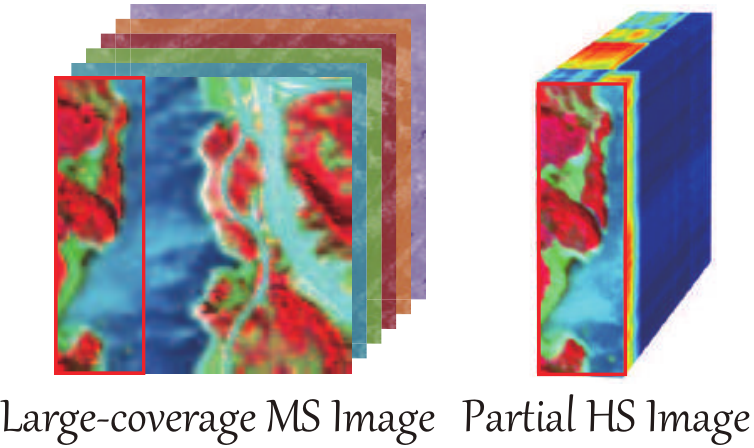}
            \label{fig:JR}
		}
         \caption{Three used HS-MS datasets for performance evaluation of SSR products in terms of reconstruction, classification, and unmixing.}
\label{fig:Convergence}
\end{figure*}

\textit{Optimization with respect to $\mathbf{J}$ and $\mathbf{K}$}: The low-rank problem can be effectively solved by the \textit{Singular Value Thresholding (SVT)} used in \cite{liu2012robust,hong2018sulora}. The proximal operator can be generalized to the following steps.
\begin{itemize}
    \item Step 1. Given a variable $\mathbf{G}$ with a $r$-rank, we first perform singular value decomposition (SVD) on $\mathbf{G}$:
    \begin{equation}
    \label{eq11}
    \begin{aligned}
       \mathrm{SVD}(\mathbf{G}):=\mathbf{U}\mathbf{S}\mathbf{V},\;\; \mathbf{S}=\mathrm{diag}(\{s_{k}\}_{1\leq k \leq r}).
    \end{aligned}
    \end{equation}
     \item Step 2. For any $\tau \geq 0$, a soft-thresholding operator, denoted as $\mathcal{D}_{r}$, is further adopted as follows:
    \begin{equation}
    \label{eq12}
    \begin{aligned}
       \mathcal{D}(\mathbf{G}):=\mathbf{U}\mathcal{D}_{r}(\mathbf{S})\mathbf{V},\;\; \mathcal{D}_{r}(\mathbf{S})=\mathrm{diag}(\{s_{k}-\tau\}^{+}).
    \end{aligned}
    \end{equation}
    \item Step 3. The nuclear norm of $\mathbf{G}$, namely $\norm{\mathbf{G}}_{*}$, can be obtained by $\norm{\mathcal{D}_{r}(\mathbf{S})}_{1,1}$.
\end{itemize}
Thereby, the update rule of $\mathbf{J}$ and $\mathbf{K}$ are given by
\begin{equation}
\label{eq13}
\begin{aligned}
       &\mathbf{J} \leftarrow \max\{\mathbf{0}, \mathcal{D}_{\beta}(\mathbf{D}_{h}-\mathbf{\Lambda}_{2}/\mu)\},\\
       &\mathbf{K} \leftarrow \max\{\mathbf{0}, \mathcal{D}_{\beta}(\mathbf{D}_{m}-\mathbf{\Lambda}_{3}/\mu)\},
\end{aligned}
\end{equation}
where $\mathcal{D}_{\beta}(\bullet)$ is defined as the general \textit{SVT} operator, i.e., $\mathbf{U}\mathbf{S}_{r}\mathbf{V}$, where $\mathbf{S}_{r}=\mathrm{diag}(\max\{0, s_{k}-\beta/\mu\})$.

\textit{Optimization with respect to $\mathbf{Z}$}: The $\ell_1$-norm of $\mathbf{Z}$ can be optimized by solving the following problem
\begin{equation}
\label{eq14}
\begin{aligned}
       \mathop{\min}_{\mathbf{Z}}\;&\beta\norm{\mathbf{Z}}_{1,1}+\mathbf{\Lambda}_{1}(\mathbf{Z}-\mathbf{X})+\frac{\mu}{2}\norm{\mathbf{Z}-\mathbf{X}}_{\F}^{2},
\end{aligned}
\end{equation}
which can be well solved using a well-known \textit{soft threshold} operator \cite{bioucas2010alternating,hong2019learnable}, that is, 
\begin{equation}
\label{eq15}
\begin{aligned}
       \mathbf{Z} \leftarrow \mathrm{sign}(\mathbf{X}-\mathbf{\Lambda}_{1}/\mu)\odot\max\{\mathbf{0}, |\mathbf{X}-\mathbf{\Lambda}_{1}/\mu|-\alpha/\mu\},
\end{aligned}
\end{equation}
where $\odot$ denotes the element-wise Schur-Hadamard product.

\textit{Optimization with respect to $\{\mathbf{\Lambda}_{i}\}_{i=1}^{3}$}: The Lagrange multipliers can be updated by
\begin{equation}
\label{eq16}
\begin{aligned}
       &\mathbf{\Lambda}_{1} \leftarrow \mathbf{\Lambda}_{1}+\mu(\mathbf{Z}-\mathbf{X}),\\
       &\mathbf{\Lambda}_{2} \leftarrow \mathbf{\Lambda}_{2}+\mu(\mathbf{J}-\mathbf{D}_{h}),\\
       &\mathbf{\Lambda}_{3} \leftarrow \mathbf{\Lambda}_{3}+\mu(\mathbf{K}-\mathbf{D}_{m}).
\end{aligned}
\end{equation}
More specific optimization procedures for \textit{D-Step} as shown in Eq. (\ref{eq3}) are detailed in \textbf{Algorithm~1}.

\subsubsection{S-Step Solver} The optimization problem is nothing but a classic sparse coding, which has been well solved by many excellent works \cite{bioucas2010alternating,mairal2010online,mairal2014sparse,kang2020learning}. The ADMM-based optimizer has been proven to be effective for a fast and accurate solution. Similarly, we introduce an additional auxiliary variable $\mathbf{O}$ into the problem (\ref{eq2}) to replace the variable $\mathbf{Y}$ in the term of $\ell_{1}$-norm, yielding the following augmented Lagrangian function
\begin{equation}
\label{eq17}
\begin{aligned}
       \mathcal{L}&(\mathbf{Y},\mathbf{O},\mathbf{\Delta})=\frac{1}{2}\norm{\mathbf{M}_{out}-\mathbf{D}_{m}\mathbf{Y}}_{\F}^{2}+\eta\norm{\mathbf{O}}_{1,1}\\
       &+\mathbf{\Delta}^\top(\mathbf{O}-\mathbf{Y})+\frac{\rho}{2}\norm{\mathbf{O}-\mathbf{Y}}_{\F}^{2}, \;\; \; \mathrm{s.t.} \; \; \mathbf{1}^\top\mathbf{Y}=\mathbf{1},
\end{aligned}
\end{equation}
where $\mathbf{\Delta}$ and $\rho$ denote the Lagrange multiplier and the Lagrange penalty parameter, respectively.

To solve the problem (\ref{eq17}), we then have 

\textit{Optimization with respect to $\mathbf{Y}$}: The sparse coefficients can be estimated by solving the constrained least squares regression as follows:
\begin{equation}
\label{eq18}
\begin{aligned}
       \mathop{\min}_{\mathbf{Y}}&\frac{1}{2}\norm{\mathbf{M}_{out}-\mathbf{D}_{m}\mathbf{Y}}_{\F}^{2}+\mathbf{\Delta}^\top(\mathbf{O}-\mathbf{Y})+\frac{\rho}{2}\norm{\mathbf{O}-\mathbf{Y}}_{\F}^{2},\\
       &\mathrm{s.t.} \; \; \mathbf{1}^\top\mathbf{Y}=\mathbf{1}.
\end{aligned}
\end{equation}
Similarly to the problem (\ref{eq5}), we have the closed-form solution of the problem (\ref{eq18}):
\begin{equation}
\label{eq19}
\begin{aligned}
       \mathbf{Y} \leftarrow \tilde{\mathbf{A}}^{-1}\tilde{\mathbf{B}}-\tilde{\mathbf{C}}(\mathbf{I}_{1\times L}\tilde{\mathbf{A}}^{-1}\tilde{\mathbf{B}}-1),
\end{aligned}
\end{equation}
where 
\begin{equation}
\nonumber
\begin{aligned}
       &\tilde{\mathbf{A}}=\mathbf{D}_{m}^\top\mathbf{D}_{m}+\rho\mathbf{I}_{L\times L},\\
       &\tilde{\mathbf{B}}=\mathbf{D}_{m}^\top\mathbf{M}_{out}+\rho\mathbf{O}+\mathbf{\Delta},\\
       &\tilde{\mathbf{C}}=\tilde{\mathbf{A}}^{-1}\mathbf{I}_{L\times 1}(\mathbf{I}_{1\times L}\tilde{\mathbf{A}}^{-1}\mathbf{I}_{L\times 1})^{-1}.
       \end{aligned}
\end{equation}

\textit{Optimization with respect to $\mathbf{O}$}: Likewise, the variable $\mathbf{O}$ can be updated by following the same rule as Eq. (\ref{eq15}) 
\begin{equation}
\label{eq20}
\begin{aligned}
       \mathbf{O} \leftarrow \mathrm{sign}(\mathbf{Y}-\mathbf{\Delta}/\rho)\odot\max\{\mathbf{0}, |\mathbf{Y}-\mathbf{\Delta}/\rho|-\eta/\rho\}.
\end{aligned}
\end{equation}

\textit{Optimization with respect to $\mathbf{\Delta}$}: By employing the same rule in Eq. (\ref{eq16}), the Lagrange multiplier in each step is written as
\begin{equation}
\label{eq21}
\begin{aligned}
       \mathbf{\Delta} \leftarrow \mathbf{\Delta}+\rho(\mathbf{O}-\mathbf{Y}).
\end{aligned}
\end{equation}
\textbf{Algorithm~2} gives more optimization details for the \textit{S-Step}.

\begin{figure*}[!t]
	  \centering
		\subfigure{
			\includegraphics[width=1\textwidth]{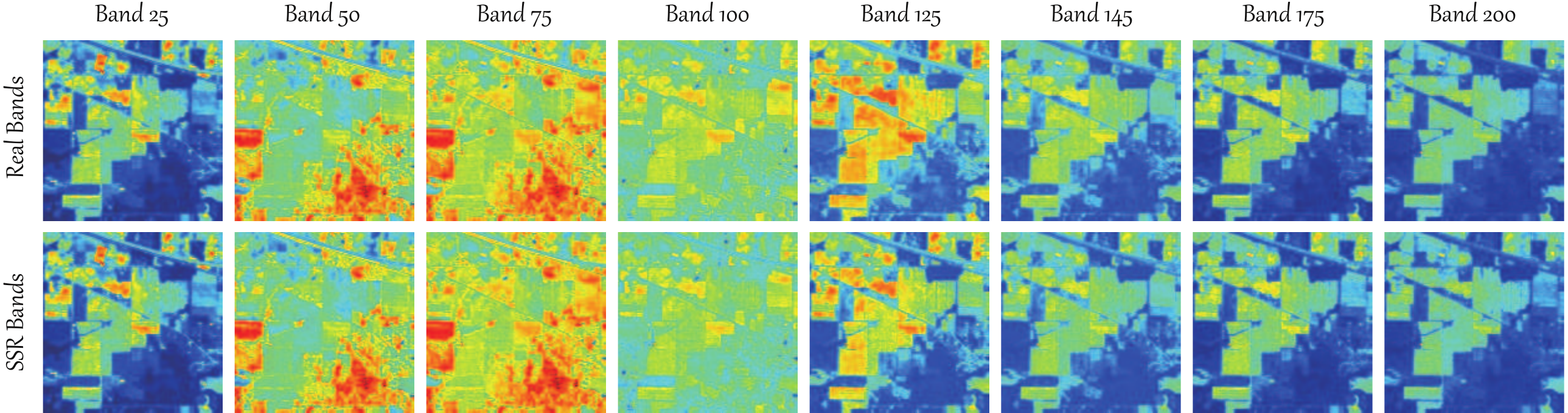}
		}
        \caption{Selective band visualization of the real HS image and the SSR product using the proposed J-SLoL model on the Indian Pines data.}
\label{fig:IP_band}
\end{figure*}

\begin{figure*}[!t]
	  \centering
		\subfigure{
			\includegraphics[width=1\textwidth]{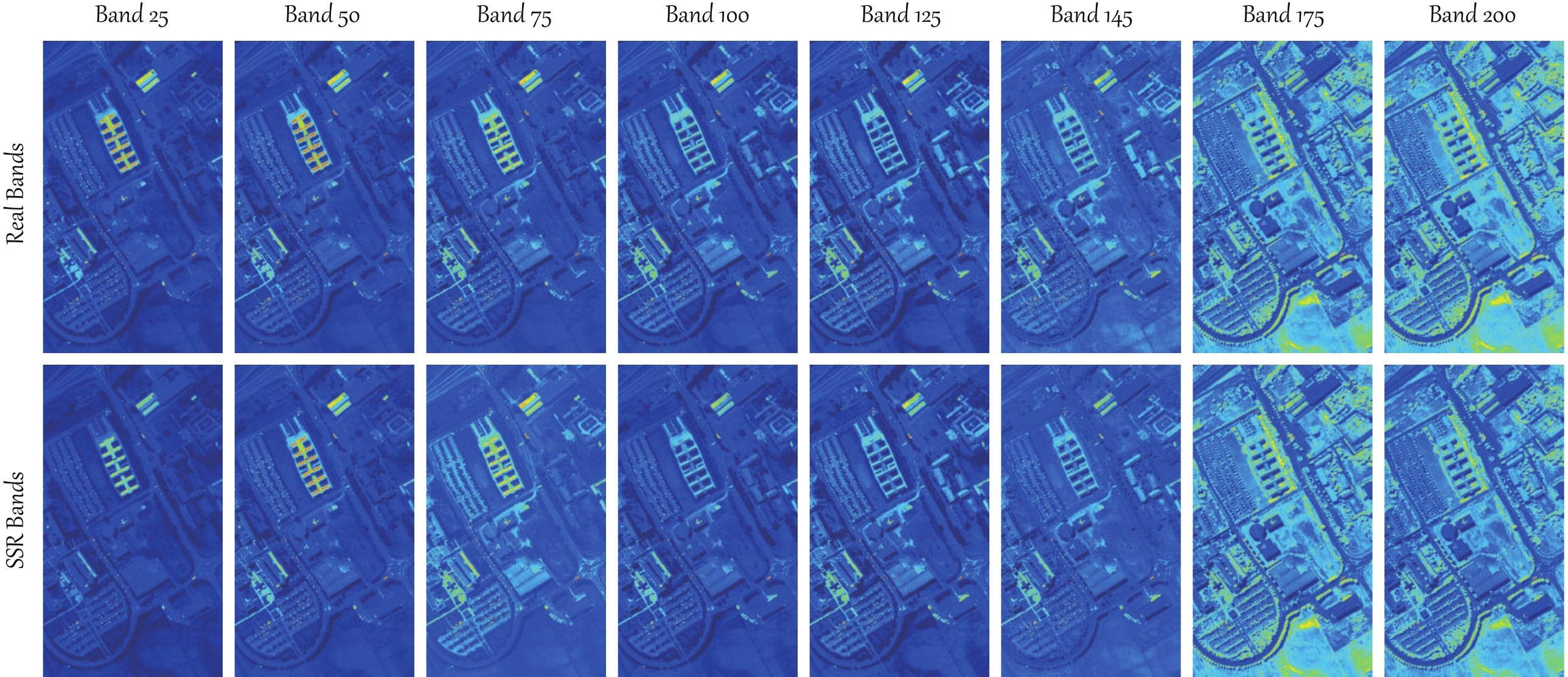}
		}
        \caption{Selective band visualization of the real HS image and the SSR product using the proposed J-SLoL model on the Pavia University data.}
\label{fig:PU_band}
\end{figure*}

\subsection{Convergence Analysis}
The solution of the optimization problems both (\ref{eq3}) and (\ref{eq17}) can be well obtained by using a ADMM solver. Actually, the used ADMM in this paper can be generalized to \emph{inexact} Augmented Lagrange Multiplier (ALM) \cite{lin2010augmented}, whose convergence has been theoretically guaranteed as long as the number of block is less than three. Although the multi-block ADMM optimization, e.g., our problem (\ref{eq3}), is still lack of a \emph{strictly} mathematical proof, yet its convergence has been well proven and maintained by tons of practical cases, such as \cite{Xu2012ADMM,Zhang2013MultiADMM,Zhou2017MultiADMM,hong2019learning}.

\section{Experiments}
We evaluate the quality of SSR product of MS images from three different perspectives.
\begin{itemize}
    \item Reconstruction. We directly measure the differences between the real HS image and the SSR product.
    \item Classification. The goal of SSR is to generate high-quality HS products for the subsequent high-level applications, e.g., classification. Classification can thus be seen as an effective tool to verify spectrally physical properties of reconstructed HS images (or SSR products).
    \item Unmixing. Material mixture is also an unique to the HS image. As a result, the evaluation of SSR products can, to a great extent, be determined by the unmixing performance. 
\end{itemize}
Moreover, three HS-MS datasets are used for the SSR task, where the first two are applied for the performance comparison in terms of reconstruction and classification, and the last one is used for unmixing evaluation.

\begin{table}[!t]
\centering
\caption{Reconstruction performance comparison of different methods in terms of five indices on the HS-MS Indian Pines dataset. The best results are shown in bold.}
\begin{tabular}{c|ccccc}
\toprule[1.5pt]
Methods & RMSE & PSNR & SAD & SSIM & ERGAS\\
\hline \hline
PwC & 0.0057 & 38.7302 & 0.0125& 0.7932 & 1.0092\\
CRISP \cite{winter2007hyperspectral} & 0.0039 & 39.3170 & 0.0092 & 0.7782 & 0.8820\\
Sun's \cite{sun2014enhancement} & 0.0039 & 41.4753 & 0.0093 & 0.7999 & 0.7110\\
Yokoya's \cite{yokoya2018spectral} & 0.0036 & 42.2450 & 0.0087 & 0.8201 & 0.6546 \\
Arad's \cite{arad2016sparse} & 0.0035 & 42.4517 & 0.0084 & 0.8344 & 0.6405\\ 
\hline 
J-SLoL & \bf 0.0032 & \bf 43.1547 & \bf 0.0080 &  \bf 0.8688 & \bf 0.6044\\
\hline \hline
Ideal Value & $\downarrow$ 0 & $\uparrow$ $\infty$  & $\downarrow$ 0 & $\uparrow$ 1 & $\downarrow$ 0\\
\bottomrule[1.5pt]
\end{tabular}
\label{tab:Recon_IP}
\end{table}

\begin{table}[!t]
\centering
\caption{Reconstruction performance comparison of different methods in terms of five indices on the HS-MS Pavia University dataset. The best results are shown in bold.}
\begin{tabular}{c|ccccc}
\toprule[1.5pt]
Methods & RMSE & PSNR & SAD & SSIM & ERGAS\\
\hline \hline
PwC & 0.0133 & 38.1921 & 0.0487 & 0.8804 & 3.4181\\
CRISP \cite{winter2007hyperspectral} & 0.0074 & 45.8438 & 0.0367 & 0.9608 & 1.9097\\
Sun's \cite{sun2014enhancement} & 0.0070 & 46.2677 & 0.0355 & 0.9584 & 1.8140\\
Yokoya's \cite{yokoya2018spectral} & 0.0068 & 46.4856 & 0.0351 & 0.9619 & 1.7705\\
Arad's \cite{arad2016sparse} & 0.0069 & 46.4749 & 0.0350 & \bf 0.9674 & 1.7782\\ 
\hline 
J-SLoL & \bf 0.0060 & \bf 47.4549 & \bf 0.0328 & 0.9618 & \bf 1.6403\\
\hline \hline
Ideal Value & $\downarrow$ 0 & $\uparrow$ $\infty$  & $\downarrow$ 0 & $\uparrow$ 1 & $\downarrow$ 0\\
\bottomrule[1.5pt]
\end{tabular}
\label{tab:Recon_PU}
\end{table}

\subsection{Dataset Description}

\subsubsection{Indian Pines Data}
This widely-used HS image was collected by the optical sensor -- Airborne Visible / Infrared Imaging Spectrometer (AVIRIS) -- over the Indiana state, USA. It consists of $145\times 145$ pixels with $220$ spectral channels. To meet the requirement of our SSR's problem setting, the corresponding MS image is simulated by using the spectral response functions (SRFs) of Sentinel-2 and a sub-image is selected with the size of $145\times 45$ as the partially overlapped region of HS and MS images, as shown in Fig. \ref{fig:IP}. Furthermore, there are $16$ to-be-investigated categories in the studied scene, and they will be used for the part of classification evaluation. More details about training and test sets can be found in \cite{hang2019cascaded}.

\subsubsection{Pavia University Data}
The second HS data was captured by the Reflective Optics System Imaging Spectrometer (ROSIS) sensor covering the Pavia University, Pavia, Italy, which was used for IEEE GRSS data fusion contest 2008 \cite{licciardi2009decision}. The image comprises $103$ spectral bands covering the wavelength range from $430nm$ to $860nm$. Also, this scene consists of $610\times 340$ pixels at a $1.3m$ GSD, including 9 classes used for the land cover classification. To make a relatively fair comparison, we adopt a set of fixed training and test samples widely used in many researches \cite{hong2020invariant}.  Similarly, the simulated MS image is generated by using the SRFs of QuickBird, yielding the size of $610\times 340\times 4$, $610\times 50$ of which are selected as the HS-MS overlapped region (see Fig. \ref{fig:PU}).

\subsubsection{Jasper Ridge Data}
The HS scene was acquired using the AVIRIS instrument over a rural area at Jasper Ridge, California, USA. A widely-used region of interest (ROI) with $100\times 100$ pixels at a GSD of $20m$ and $198$ spectral bands in the range of $380nm$ to $2500nm$ is used in our experiments. A Sentinel-2 MS product is simulated using SRFs on the HS image, and there is the size of $100\times 30$ pixels in the overlapped part between HS and MS images, as shown in Fig. \ref{fig:JR}. Four main materials, such as \textit{\#1 Tree}, \textit{\#2 Water}, \textit{\#3 Soil}, and \textit{\#4 Road}, are involved in this scene with ground truth of abundance maps given from the website\footnote{https://rslab.ut.ac.ir/data}.

\begin{figure*}[!t]
	  \centering
		\subfigure{
			\includegraphics[width=1\textwidth]{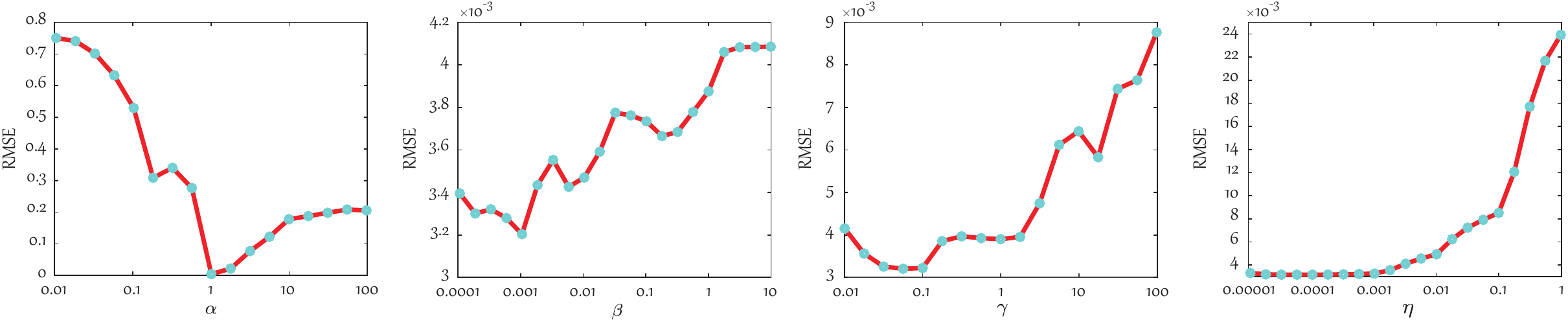}
		}
        \caption{Parameter sensitivity analysis of the proposed J-SLoL model in terms of four regularization parameters, e.g., $\alpha$, $\beta$, $\gamma$ in Eq. (\ref{eq1}), and $\eta$ in Eq. (\ref{eq2}) on the Indian Pines data.}
\label{fig:para}
\end{figure*}

\begin{table}[!t]
\centering
\caption{Classification performance comparison of different methods in terms of OA (\%), AA (\%), and $\kappa$ (\%) as well as the accuracy for each class on the Indian Pines dataset. R-HS means the real HS image. The best results are shown in bold.}
\resizebox{0.5\textwidth}{!}{
\begin{tabular}{c|ccccc|c||c}
\toprule[1.5pt]
 & PwC & CRISP & Sun's & Yokoya's & Arad's & J-SLoL & R-HS\\
\hline \hline
OA& 52.92 & 63.74 & 60.77 & 61.60 & 64.67 & \bf 65.14 & 65.89\\
AA & 65.34	& 73.04 & 71.15 & 71.90 & \bf 76.27 & 73.37 & 75.71 \\
$\kappa$ & 47.31 & 59.00 & 55.69 & 56.62 & 60.10 & \bf 60.53 & 61.48\\
\hline
C1 & 35.55 & 44.94 & 43.28 & 46.10 & 43.64 & \bf 48.55 & 51.66\\
C2 & 42.47 & 48.34 & 43.75 & 43.88 & \bf 50.38 & 48.60 & 57.40\\
C3 & 65.76	& 63.59 & 61.96 & 61.96 & \bf 69.57 & 67.39 & 70.65\\
C4 & 68.46	& \bf 88.14 & 74.72 & 74.94 & 86.80 & 86.13 & 88.14\\
C5 & 80.63	& 81.78 & 82.93 & 83.21 & \bf 83.93 & 83.36 & 81.78\\
C6 & 78.82	& \bf 93.62 & 90.89 & \bf 93.62 & 92.71 & 93.39 & 95.90\\
C7 & 48.69	& 71.24 & 66.45 & 67.10 & \bf 78.65 & 73.42 & 66.56\\
C8 & 43.26	& 53.14 & 51.08 & 50.74 & 54.76 & \bf 56.12 & 55.21\\
C9 & 44.5 & 48.94 & 42.73 & 44.15 & \bf 50.53 & 46.10 & 53.01\\
C10 & 94.44 & \bf 97.53 & \bf 97.53 & \bf 97.53 & 96.30 & \bf 97.53 & 98.15\\
C11 & 68.17 & 83.28 & 82.23 & 83.68 & 78.94 & \bf 83.92 & 82.88\\
C12 & 39.70 & 52.42 & 47.27 & 47.27 & \bf 55.76 & 50.30 & 50.91\\
C13 & 95.56 & 97.78 & \bf 100 & \bf 100 & 97.78 & 97.78 & 97.78\\
C14 & 66.67 & 82.05 & 71.79 & 74.36 & \bf 89.74 & 79.49 & 79.49\\
C15 & 72.73 & 81.82 & 81.82 & 81.82 & \bf 90.91 & 81.82 & 81.82\\
C16 & \bf 100 & 80.00 & \bf 100 & \bf 100 & \bf 100 & 80.00 & 100\\
\bottomrule[1.5pt]
\end{tabular}
}
\label{tab:Class_IP}
\end{table}

\subsection{Reconstruction-based Evaluation}
Several important indices from the perspective of reconstruction, e.g., root mean square error (RMSE), peak signal to noise ratio (PSNR), spectral angle distance (SAD), structural similarity index (SSIM), and erreur relative globale adimensionnelle de synth{\`e}se (ERGAS), are employed to quantitatively evaluate the performance of SSR of MS images. In addition, we select four state-of-the-art baselines related to the SSR task, including pixel-wise copy (PwC)\footnote{We directly copy the HS pixel from the overlapped region into the unknown HS pixel, according to the similarity measurement in the MS image.}, color resolution improvement software package (CRISP) \cite{winter2007hyperspectral}, Sun's \cite{sun2014enhancement}, Yokoya's \cite{yokoya2018spectral}, and Arad's \cite{arad2016sparse}, in comparison with our J-SLoL model.

Table \ref{tab:Recon_IP} lists the quantitative results of the aforementioned compared algorithms in five indices on the Indian Pines data. Overall, the PwC approach yields the poor reconstruction performance in nearly all indices, except the SSIM value that is slightly higher than CRISP's. By making use of grouping strategy, the spectrally-enhanced performance of Sun's algorithm is further improved  compared to the original CRISP, specifically in PSNR and ERGAS. Inspired by the current success and good theoretical support in sparse representation, Yokoya's and Arad's methods show great potential in the SSR task, yielding a moderate improvement in all measures. Please note that the main difference between Yokoya's and Arad's methods lies in the MS dictionary construction. The former directly takes the overlapped MS data as the dictionary, while the latter generates the MS dictionary by performing the linear interpolation on HS data. As a result, the MS dictionary obtained by Arad's method might be more correlated with the HS's, yielding relatively higher reconstruction results compared to Yokoya's. Remarkably, the proposed J-SLoL model performs better than others at a comprehensive increase of around $0.0004$ RMSE, $1$ PSNR, $0.0007$ SAD, $0.04$ SSIM, and $0.05$ ERGAS, compared to the second best approach. Furthermore, Fig. \ref{fig:IP_band} visualizes several selected bands, where there is a very small visual difference between the real HS image and reconstructed one obtained by J-SLoL. This, to a great extent, demonstrates the effectiveness of the proposed method.

\begin{table}[!t]
\centering
\caption{Classification performance comparison of different methods in terms of OA (\%), AA (\%), and $\kappa$ (\%) as well as the accuracy for each class on the Pavia University dataset. R-HS means the real HS image. The best results are shown in bold.}
\resizebox{0.5\textwidth}{!}{
\begin{tabular}{c|ccccc|c||c}
\toprule[1.5pt]
 & PwC & CRISP & Sun's  & Yokoya's & Arad's & J-SLoL & R-HS\\
\hline \hline
OA & 66.29	& 70.86 & 70.68 & 70.50 & 70.39 & \bf 71.15 & 71.85\\
AA & 76.09	& 80.41 & 80.31 & 80.35 & 80.25 & \bf 80.48 & 81.15\\
$\kappa$ & 58.53 & 63.85 & 63.60 & 63.43 & 63.30 & \bf 64.16 & 65.01\\
\hline
C1 & 71.56	& 73.88 & 73.56 & 73.77 & \bf 73.91 & 73.64 & 73.17\\
C2 & 53.91	& 59.66 & 59.62 & 59.00 & 58.79 & \bf 60.42 & 61.32\\
C3 & 55.45	& \bf 56.88 & \bf 56.88 & 56.69 & 56.41 & 56.74 & 60.17\\
C4 & 96.64	& 97.39 & 97.42 & \bf 97.45 & 97.42 & 97.29 & 97.39\\
C5 & 98.88	& \bf 99.18 & 99.03 & 99.11 & 99.11 & 99.11 & 99.26\\
C6 & 65.52	& 71.11 & 69.78 & 70.25 & 70.17 & \bf 71.13 & 73.35\\
C7 & 71.05	& 83.53 & 84.51 & \bf 84.59 & 84.36 & 83.98 & 84.96\\
C8 & 81.75	& 86.37 & \bf 86.61 & 86.58 & 86.50 & 86.42 & 85.36\\
C9 & 90.07	& \bf 95.67 & 95.35 & \bf 95.67 & 95.56 & 95.56 & 95.35\\
\bottomrule[1.5pt]
\end{tabular}
}
\label{tab:Class_PU}
\end{table}

\begin{figure*}[!t]
	  \centering
		\subfigure{
			\includegraphics[width=1\textwidth]{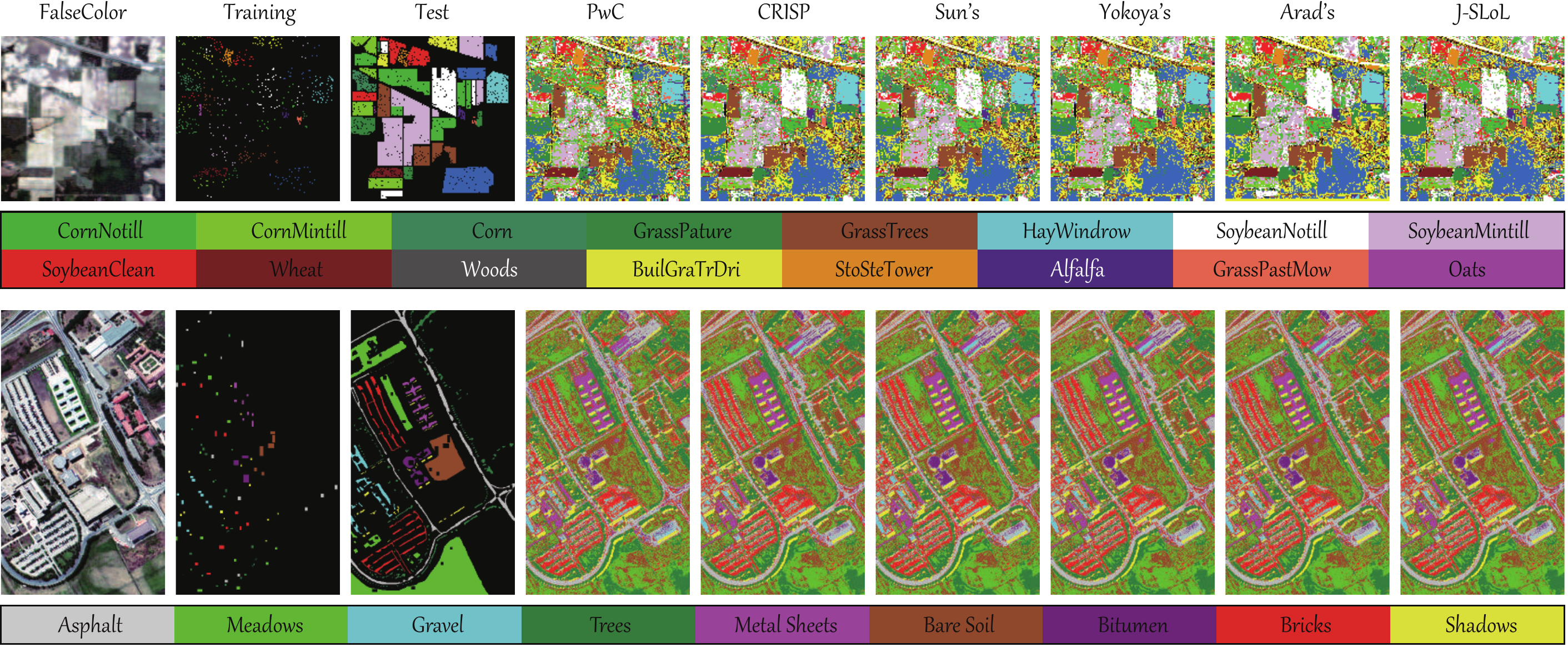}
		}
        \caption{Visualization of false-color images, training and test sample distribution, and classification maps by using different SSR algorithms on the two datasets: Indian Pines (top) and Pavia University (bottom).}
\label{fig:CM}
\end{figure*}

Similarly, there is a basically identical trend between the Indian Pines data and the Pavia University data, in either quantitative (see Table \ref{tab:Recon_PU}) or qualitative results (see Fig. \ref{fig:PU_band}). The main difference lies in that the second datasets are more challenging due to the larger image size and less spectral bands, leading to the limitations in reconstructing detailed information, e.g., texture. This can be well explained by different indices. Compared to those in the Indian Pines data, RMSE, SAD, and ERGAS are relatively low while PSNR and SSIM reflecting the image structural information are higher.

\noindent \textbf{Parameter Sensitivity Analysis.} As the quality of the SSR product is closely associated with the parameter setting in our J-SLoL model, i.e., $\alpha$, $\beta$, $\gamma$ in \textit{D-Step} and $\eta$ in \textit{S-Step}, hence the performance gain in terms of RMSE is investigated by changing these parameters in a proper range on the Indian Pines data. We can see from the Fig. \ref{fig:para} that the parameter $\alpha$ plays a dominant role in dictionary learning, while other parameters in \textit{D-Step} and \textit{S-Step} have also important effects on the whole SSR process. In detail, the optimal parameter combination  $(\alpha,\beta,\gamma,\eta)$ can be given as $(1,0.001,0.1,0.0001)$ as shown in Fig. \ref{fig:para}. We also found that this set of parameter setting is relatively stable, we therefore apply them in the rest of datasets for simplicity. 

\subsection{Classification-based Evaluation}
Classification, as a potential high-level application, has been proven to be effective for model performance assessment, where three common indices: overall accuracy (OA), average accuracy (AA), and kappa coefficient ($\kappa$) are used in our experiments. Note that a simple nearest neighbor (NN) classifier is applied for the classification task. This is because if more advanced classifiers are used, we might confuse the additional performance gain from these classifiers or our SSR products.

Tables \ref{tab:Class_IP} and \ref{tab:Class_PU} list the quantitative comparison between the real HS image and different SSR algorithms in terms of OA, AA, and $\kappa$ as well as the accuracy for each class on the two same datasets, and Fig. \ref{fig:CM} shows the corresponding classification maps. Due to only pixel-based copy operation, the PwC method fails to reconstruct high-quality HS data well, yielding poor classification performance on both datasets. Conversely, the J-SLoL model as expected outperforms other competitors, despite only a slight improvement in classification accuracies. It should be noted, however, that the results of our J-SLoL method are very close to those using the real HS image. This might directly demonstrate the superiority of the proposed strategy. Furthermore, there is also a similar trend in the visual comparison in terms of classification maps (\textit{cf.} Figs. \ref{fig:CM}). Intuitively, the classification maps of the proposed J-SLoL model are more similar to those obtained by the real HS image from either structural information or detailed textures. Moreover, our approach is capable of making the classification maps relatively smooth in certain classes, such as \textit{HayWindrow} and \textit{SoybeanNotill} in the first data, and \textit{Soil} in the second one.

\begin{table*}[!t]
\centering
\caption{Reconstruction and unmixing performance comparison of different methods on the Jasper Ridge dataset. R-HS means the real HS image. The best results are shown in bold.}
\begin{tabular}{c|ccccc|ccc}
\toprule[1.5pt]
\multirow{2}{*}{Model} & \multicolumn{5}{c|}{Reconstruction} & \multicolumn{3}{c}{Unmixing}\\
\cline{2-9} & RMSE & PSNR & SAD & SSIM & ERGAS & aRMSE & rRMSE & aSAM\\
\hline \hline
PwC & 0.0370 & 28.1639 & 0.0772 & 0.7508 & 8.2955 & 0.1802 $\pm$ 0.0147 & 0.0234 $\pm$ 0.0069 & 0.1224 $\pm$ 0.0233\\
CRISP & 0.0320 & 35.6180 & 0.0764 & 0.9304 & 6.8960 & 0.1789 $\pm$ 0.0118 & \bf 0.0222 $\pm$ 0.0070 & 0.1239 $\pm$ 0.0279\\
Sun's & 0.0298 & 36.1892 & 0.0598 & 0.9204 & 6.7216 & 0.1767 $\pm$ 0.0139 & 0.0238 $\pm$ 0.0068 & 0.1178 $\pm$ 0.0243\\
Yokoya's & 0.0292 & 36.3338 & 0.0590 & 0.9219 & 6.5876 & 0.1764 $\pm$ 0.0133 & 0.0235 $\pm$ 0.0067 & 0.1171 $\pm$ 0.0243\\
Arad's & 0.0282 & 36.5349 & 0.0578 & \bf 0.9320 & 6.3419 & 0.1763 $\pm$ 0.0132 & 0.0232 $\pm$ 0.0067 & 0.1162 $\pm$ 0.0243\\
\hline
J-SLoL & \bf 0.0271 & \bf 36.7630 & \bf 0.0565 & 0.9311 & \bf 6.0605 & \bf 0.1762 $\pm$ 0.0135 & 0.0229 $\pm$ 0.0068 & \bf 0.1152 $\pm$ 0.0244\\
\hline \hline
R-HS & 0 & $\infty$ & 0 & 1 & 0 & 0.1760 $\pm$ 0.0126 & 0.0206 $\pm$ 0.0061 & 0.1175 $\pm$ 0.0215\\
\bottomrule[1.5pt]
\end{tabular}
\label{tab:Class_UM}
\end{table*}

\begin{figure*}[!t]
	  \centering
		\subfigure{
			\includegraphics[width=1\textwidth]{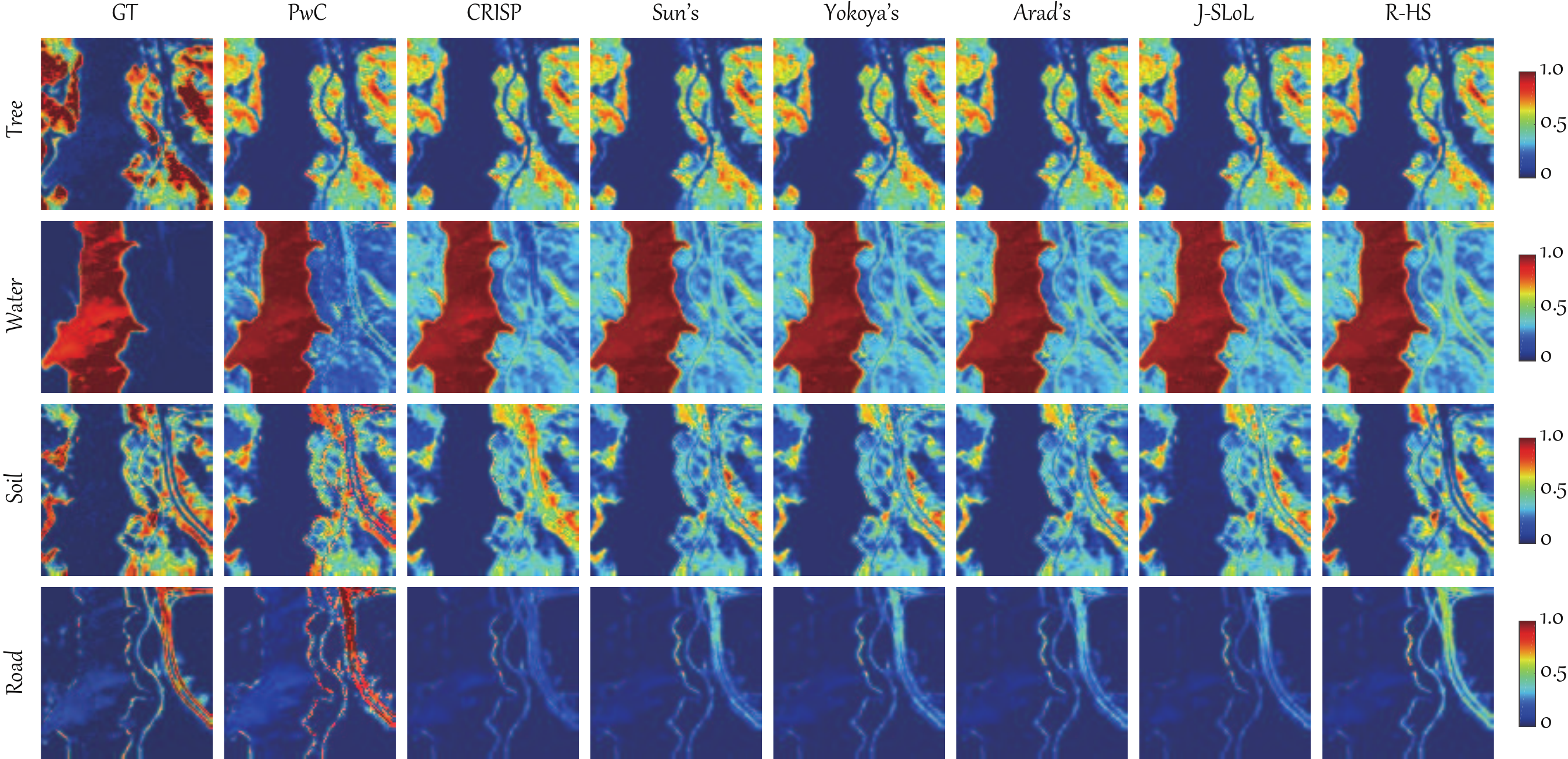}
		}
        \caption{Abundance maps of four materials using different SSR algorithms and the real HS image.}
\label{fig:AB}
\end{figure*}

\subsection{Unmixing-based Evaluation}
Due to the low spatial resolution, multiple materials are highly mixed within one pixel in the HS image. As a result, spectral unmixing can be regarded as a feasible solution for quality assessment of SSR of MS images. More specifically, the fully constrained least squares unmixing (FCLSU) \cite{heinz2001fully} algorithm with three popular criteria \cite{hong2019augmented}, including abundance overall root mean square error (aRMSE), reconstruction overall root mean square error (rRMSE), and average spectral angle mapper (aSAM), is used to quantify the unmixing performance in the following experiments. 

Reconstruction and unmixing are successively conducted to quantitatively evaluated the algorithm performance, as listed in Table \ref{tab:Class_UM}. Roughly, there is a consistent trend in performance gain from PwC to Arad's in the Jasper Ridge data. The PwC holds relatively bad performance in both reconstruction and unmixing. For the CRISP approach, it brings a dramatic  improvement on the basis of the PwC, while its modified model, i.e., Sun's, obtains better results. Inspired by the sparsity-promoting assumption, Yokoya's algorithm achieves a competitive performance. Similarly, Arad's method performs slightly better than Yokoya's, possibly owing to the use of high-quality spectral dictionary. Not unexpectedly, our proposed J-SLoL observably exceeds other compared methods, particularly in several important indices, such as RMSE and SSIM in the reconstruction task, and aRMSE in the unmixing task. Additionally, a direct proof is given by the fact that the unmixing results using the SSR product from J-SLoL are comparable to those using the real HS image under all three measures, showing the great potential of the proposed method. 

We also make a visual comparison in terms of abundance maps, as illustrated in Fig. \ref{fig:AB}. From the figure, we can see that the abundance maps estimated by PwC and CRISP are more different from those of R-HS, particularly the materials of \textit{Water}, \textit{Soil}, and \textit{Road}. By contrary, the later three methods have better visual effect, in which our J-SLoL performs more similar abundance maps, e.g., \textit{Water} and \textit{Soil}. Despite a big difference between J-SLoL and GT, this might result from the limitations of the FCLSU algorithm itself. In summary, both visual and numerical unmixing evaluation can also demonstrate that the J-SLoL is well applicable to the SSR task.

\subsection{Comparison of Computational Cost}
The computational cost of our J-SLoL model in Eqs. (\ref{eq1}) and (\ref{eq2}) is dominated by matrix products. More specifically, the update of $\mathbf{X}$, $\mathbf{D}_h$, and $\mathbf{D}_m$ in \textit{D-Step} consists of matrix multiplications and matrix inversions, yielding complexity with $\mathcal{O}(L^3+L^2N+LNP)$, $\mathcal{O}(L^3+L^2N+LNP)$, and $\mathcal{O}(L^3+L^2N+LNQ)$, respectively, while updating $\mathbf{J}$ and $\mathbf{K}$ both require computing SVDs with the orders of cost as $\mathcal{O}(\min(P^2L,PL^2))$ and $\mathcal{O}(\min(Q^2L,QL^2))$. Therefore, optimizing the variables $\mathbf{X}$ and $\mathbf{D}_{h}$ are the most expensive computational cost steps in problem (\ref{eq4}), yielding a dominant complexity $\mathcal{O}(L^3+L^2N+LNP)$ with respect to \textbf{Algorithm 1}. For \textit{S-Step}, the main per-iteration cost of \textbf{Algorithm 2} lies in the update of $\mathbf{Y}$, being similar to the update of $\mathbf{X}$, leading to a $\mathcal{O}(L^3+L^2N_1+LN_1P)$ complexity.

\begin{table}[!t]
\centering
\caption{Comparison of computational cost for different algorithms.}
\begin{tabular}{c|c}
\toprule[1.5pt]
Methods & Computational Cost\\
\hline
PwC & $\mathcal{O}(N_1NP)$\\
CRISP & $\mathcal{O}(NPQ)$\\
Sun's & $\mathcal{O}(NPQ+KPQ^2)$\\
Yokoya's & $\mathcal{O}(L^{3}+L^{2}N_{1}+LN_{1}P)$\\
Arad's & $\mathcal{O}(L^{3}+L^{2}N_{1}+LN_{1}P)$\\
J-SLoL & $\mathcal{O}(L^3+L^2N+LNP+LN_1P)$\\
\bottomrule[1.5pt]
\end{tabular}
\label{tab:CC}
\end{table}

To demonstrate the efficiency and effectiveness of the proposed J-SLoL model, we make an approximated comparison in computational cost. As listed in Table \ref{tab:CC}, PwC holds a very high cost of $\mathcal{O}(N_1NP)$ due to its pixel-to-pixel matching operation. The complexity in CRISP lies in the estimation of the regression matrix between overlapped HS and MS images, yielding a $\mathcal{O}(NPQ)$ computational cost. Sun's method involves an additional spectral matching cost behind CRISP, yielding a complexity of $\mathcal{O}(KPQ^2)$, where $K$ is a predefined number of materials. In fact, the Yokoya's and Arad's methods can be approximately seen as our \textit{S-Step}, hence its computational cost is $\mathcal{O}(L^{3}+L^{2}N_{1}+LN_{1}P)$. Although our J-SLoL performs relatively higher than Yokoya's and Arad's methods, yet the overall computational cost is acceptable, since the number of HS-MS samples in the overlapped region, i.e., $N$, are limited.

\section{Conclusion}
In this paper, a novel and promising topic -- SSR -- is introduced to enhance the spectral resolution of MS images, which has a great potential as a better alternative of the classic HS and MS fusion task (or HS-SR) with only need of partially HS data. Inspired by the effectiveness and recent success of sparse representation, we propose an effective MS enhancement algorithm, called J-SLoL, by jointly learning a low-rank dictionary pair from the overlapped HS and MS region and further inferring the unknown HS image by sharing the sparse coefficients estimated by using MS data. Beyond previous models, the proposed J-SLoL is capable of fully making use of the correspondences between HS and MS images to learn more completed HS and MS dictionaries, further yielding a more accurate HS recovery. We have to admit, however, that the linearized sparse technique remains limited in data representation and fitting, especially in large-scale and complex cases. In the future work, we would like to develop more advanced reconstruction and recovery strategies by the means of nonlinear techniques, e.g., deep learning, or by introducing the new data source, e.g., LiDAR, SAR, to further improve the model's generalization ability.

\bibliographystyle{IEEEbib}
\bibliography{HDF_ref}

\begin{IEEEbiography}[{\includegraphics[width=1in,height=1.25in,clip,keepaspectratio]{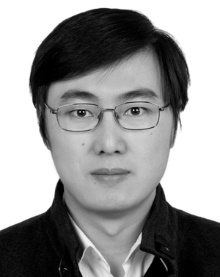}}]{Lianru Gao} (M'12-SM'18) received the B.S. degree in civil engineering from Tsinghua University, Beijing, China, in 2002, and the Ph.D. degree in cartography and geographic information system from Institute of Remote Sensing Applications, Chinese Academy of Sciences (CAS), Beijing, China, in 2007.

He is currently a Professor with the Key Laboratory of Digital Earth Science, Aerospace Information Research Institute, CAS. He also has been a visiting scholar at the University of Extremadura, Cáceres, Spain, in 2014, and at the Mississippi State University (MSU), Starkville, USA, in 2016. His research focuses on hyperspectral image processing and information extraction. In last ten years, he was the PI of 10 scientific research projects at national and ministerial levels, including projects by the National Natural Science Foundation of China (2010-2012, 2016-2019, 2018-2020), and by the Key Research Program of the CAS (2013-2015) et al. He has published more than 160 peer-reviewed papers, and there are more than 80 journal papers included by Science Citation Index (SCI). He was coauthor of an academic book entitled ``Hyperspectral Image Classification And Target Detection''. He obtained 28 National Invention Patents in China. He was awarded the Outstanding Science and Technology Achievement Prize of the CAS in 2016, and was supported by the China National Science Fund for Excellent Young Scholars in 2017, and won the Second Prize of The State Scientific and Technological Progress Award in 2018. He received the recognition of the Best Reviewers of the IEEE Journal of Selected Topics in Applied Earth Observations and Remote Sensing in 2015, and the Best Reviewers of the IEEE Transactions on Geoscience and Remote Sensing in 2017.
\end{IEEEbiography}

\vskip -2\baselineskip plus -1fil
\begin{IEEEbiography}[{\includegraphics[width=1in,height=1.25in,clip,keepaspectratio]{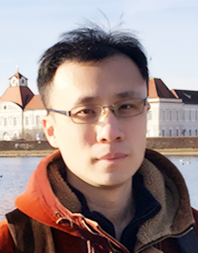}}]{Danfeng Hong}
(S'16-M'19) received the M.Sc. degree (summa cum laude) in computer vision, College of Information Engineering, Qingdao University, Qingdao, China, in 2015, the Dr. -Ing degree (summa cum laude) in Signal Processing in Earth Observation (SiPEO), Technical University of Munich (TUM), Munich, Germany, in 2019. 

Since 2015, he also worked as a Research Associate at the Remote Sensing Technology Institute (IMF), German Aerospace Center (DLR), Oberpfaffenhofen, Germany. Currently, he is research scientist and leads a Spectral Vision group at EO Data Science, IMF, DLR, and also an adjunct scientist in GIPSA-lab, Grenoble INP, CNRS, Univ. Grenoble Alpes, Grenoble, France.

His research interests include signal / image processing and analysis, pattern recognition, machine / deep learning and their applications in Earth Vision.
\end{IEEEbiography}

\vskip -2\baselineskip plus -1fil
\begin{IEEEbiography}[{\includegraphics[width=1in,height=1.25in,clip,keepaspectratio]{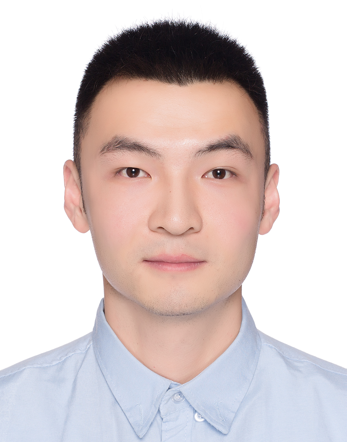}}]{Jing Yao} received the B.Sc. degree from Northwest University, Xi’an, China, in 2014. He is currently pursuing the Ph.D. degree with the School of Mathematics and Statistics, Xi’an Jiaotong University, Xi’an, China. 

From 2019 to 2020, he is a visiting student in Signal Processing in Earth Observation (SiPEO), Technical University of Munich (TUM), Munich, Germany, and at the Remote Sensing Technology Institute (IMF), German Aerospace Center (DLR), Oberpfaffenhofen, Germany.

His research interests include low-rank modeling, hyperspectral image analysis and deep learning-based image processing methods.
\end{IEEEbiography}

\vskip -2\baselineskip plus -1fil
\begin{IEEEbiography}[{\includegraphics[width=1in,height=1.25in,clip,keepaspectratio]{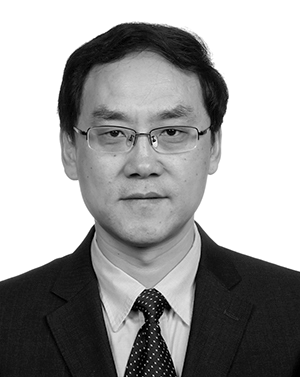}}]{Bing Zhang} (M'11–SM'12-F'19) received the B.S. degree in geography from Peking University, Beijing, China, in 1991, and the M.S. and Ph.D. degrees in remote sensing from the Institute of Remote Sensing Applications, Chinese Academy of Sciences (CAS), Beijing, China, in 1994 and 2003, respectively.

Currently, he is a Full Professor and the Deputy Director of the Aerospace Information Research Institute, CAS, where he has been leading lots of key scientific projects in the area of hyperspectral remote sensing for more than 25 years. His research interests include the development of Mathematical and Physical models and image processing software for the analysis of hyperspectral remote sensing data in many different areas. He has developed 5 software systems in the image processing and applications. His creative achievements were rewarded 10 important prizes from Chinese government, and special government allowances of the Chinese State Council. He was awarded the National Science Foundation for Distinguished Young Scholars of China in 2013, and was awarded the 2016 Outstanding Science and Technology Achievement Prize of the Chinese Academy of Sciences, the highest level of Awards for the CAS scholars.

Dr. Zhang has authored more than 300 publications, including more than 170 journal papers. He has edited 6 books/contributed book chapters on hyperspectral image processing and subsequent applications. He is the IEEE fellow and currently serving as the Associate Editor for IEEE Journal of Selected Topics in Applied Earth Observations and Remote Sensing. He has been serving as Technical Committee Member of IEEE Workshop on Hyperspectral Image and Signal Processing since 2011, and as the president of hyperspectral remote sensing committee of China National Committee of International Society for Digital Earth since 2012, and as the Standing Director of Chinese Society of Space Research (CSSR) since 2016. He is the Student Paper Competition Committee member in IGARSS from 2015-2019.
\end{IEEEbiography}

\vskip -2\baselineskip plus -1fil
\begin{IEEEbiography}[{\includegraphics[width=1in,height=1.25in,clip,keepaspectratio]{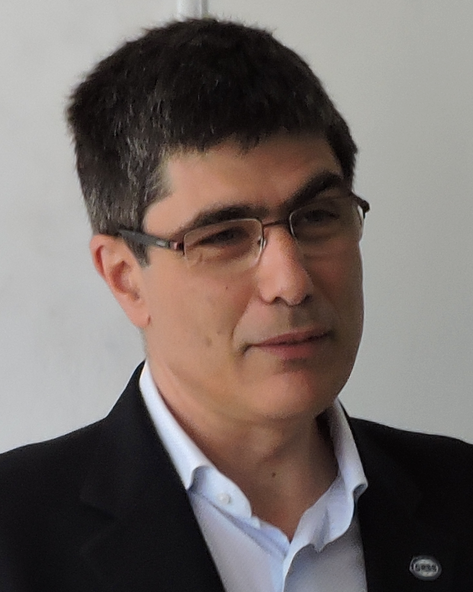}}]{Paolo Gamba} (SM'00-F'13) received the Laurea degree in Electronic Engineering ``cum laude'' from the University of Pavia, Italy, in 1989, and the Ph.D. in Electronic Engineering from the same University in 1993. 

He currently is a Professor with the University of Pavia, Italy, where he leads the Telecommunications and Remote Sensing Laboratory. He served as Editor-in-Chief of the IEEE Geoscience and Remote Sensing Letters from 2009 to 2013, and as Chair of the Data Fusion Committee of the IEEE Geoscience and Remote Sensing Society (GRSS) from October 2005 to May 2009. He has been elected in the GRSS AdCom since 2014, and served as GRSS President from 2019 to 2020. He has been the organizer and Technical Chair of the biennial GRSS / ISPRS Joint Workshops on ``Remote Sensing and Data Fusion over Urban Areas'' from 2001 to 2015. He also served as Technical Co-Chair of the 2010, 2015 and 2020 IGARSS conferences, in Honolulu (Hawaii), Milan (Italy), and Waikoloa (Hawaii), respectively.

He has been invited to give keynote lectures and tutorials in several occasions about urban remote sensing, data fusion, EO data for physical exposure and risk management. He published more than 140 papers in international peer-review journals and presented nearly 300 research works in workshops and conferences.

\end{IEEEbiography}

\vskip -2\baselineskip plus -1fil
\begin{IEEEbiography}[{\includegraphics[width=1in,height=1.25in,clip,keepaspectratio]{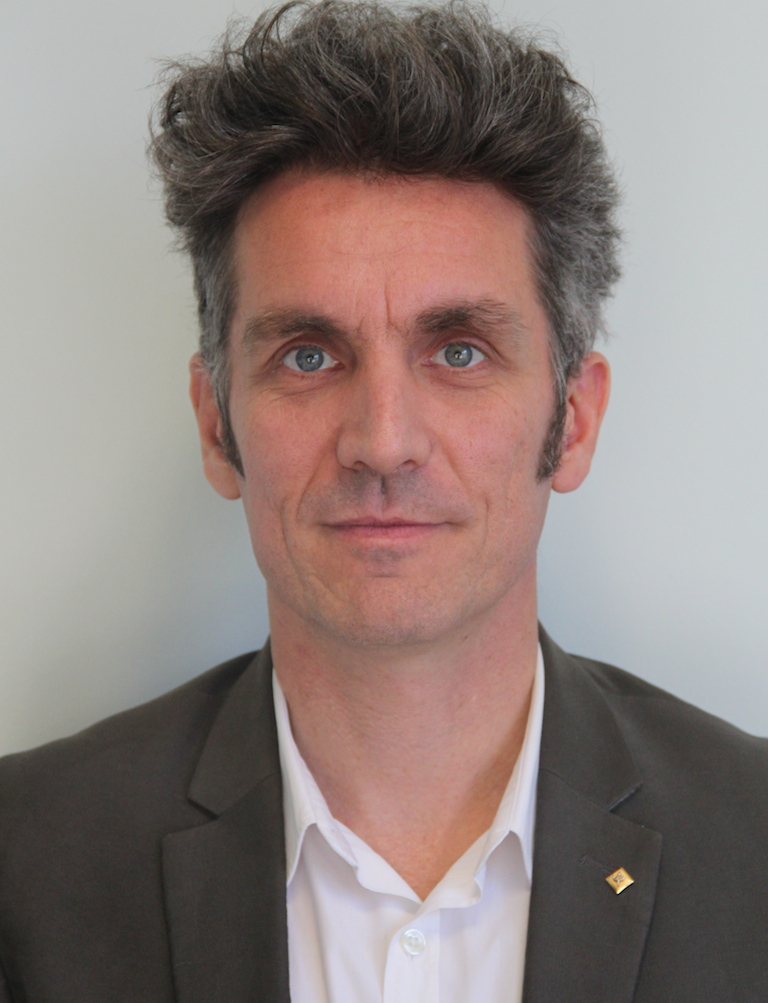}}]{Jocelyn Chanussot}
(M'04–SM'04-F'12) received the M.Sc. degree in electrical engineering from the Grenoble Institute of Technology (Grenoble INP), Grenoble, France, in 1995, and the Ph.D. degree from the Université de Savoie, Annecy, France, in 1998. Since 1999, he has been with Grenoble INP, where he is currently a Professor of signal and image processing. His research interests include image analysis, hyperspectral remote sensing, data fusion, machine learning and artificial intelligence. He has been a visiting scholar at Stanford University (USA), KTH (Sweden) and NUS (Singapore). Since 2013, he is an Adjunct Professor of the University of Iceland. In 2015-2017, he was a visiting professor at the University of California, Los Angeles (UCLA). He holds the AXA chair in remote sensing and is an Adjunct professor at the Chinese Academy of Sciences, Aerospace Information research Institute, Beijing.

Dr. Chanussot is the founding President of IEEE Geoscience and Remote Sensing French chapter (2007-2010) which received the 2010 IEEE GRS-S Chapter Excellence Award. He has received multiple outstanding paper awards. He was the Vice-President of the IEEE Geoscience and Remote Sensing Society, in charge of meetings and symposia (2017-2019). He was the General Chair of the first IEEE GRSS Workshop on Hyperspectral Image and Signal Processing, Evolution in Remote sensing (WHISPERS). He was the Chair (2009-2011) and  Cochair of the GRS Data Fusion Technical Committee (2005-2008). He was a member of the Machine Learning for Signal Processing Technical Committee of the IEEE Signal Processing Society (2006-2008) and the Program Chair of the IEEE International Workshop on Machine Learning for Signal Processing (2009). He is an Associate Editor for the IEEE Transactions on Geoscience and Remote Sensing, the IEEE Transactions on Image Processing and the Proceedings of the IEEE. He was the Editor-in-Chief of the IEEE Journal of Selected Topics in Applied Earth Observations and Remote Sensing (2011-2015). In 2014 he served as a Guest Editor for the IEEE Signal Processing Magazine. He is a Fellow of the IEEE, a member of the Institut Universitaire de France (2012-2017) and a Highly Cited Researcher (Clarivate Analytics/Thomson Reuters, 2018-2019).

\end{IEEEbiography}
\end{document}